\documentclass[12pt,preprint]{aastex}

\usepackage{amsmath, amsthm, amssymb}
\usepackage{natbib}
\usepackage{epsfig}
\usepackage[update,prepend]{epstopdf}

\usepackage[backref,breaklinks,colorlinks,citecolor=blue]{hyperref} 
\usepackage[all]{hypcap} 

\shorttitle{Evidence for superrotation in three {\it  Kepler} hot Jupiters}
\shortauthors{Faigler & Mazeh}

\begin{document}
\title{BEER analysis of {\it Kepler} and CoRoT light curves: \\
II. Evidence for superrotation in the phase curves of three {\it Kepler} hot Jupiters}

\author{S. Faigler and  T. Mazeh        }
\affil{ School of Physics and Astronomy, Raymond and Beverly Sackler Faculty of Exact Sciences, Tel Aviv University, Tel Aviv  69978, Israel}

\begin{abstract}
We analyzed the {\it Kepler} light curves of four transiting hot Jupiter systems --- KOI-13, HAT-P-7, TrES-2, and Kepler-76, which show BEaming, Ellipsoidal and Reflection (BEER) phase modulations.
The mass of the four planets can be estimated from either the beaming or the ellipsoidal amplitude, given the mass and radius of their parent stars. 
For KOI-13, HAT-P-7, and Kepler-76 we find that the beaming-based planetary mass estimate is larger than the mass estimated from the ellipsoidal amplitude, consistent with previous studies.
This apparent discrepancy may be explained by equatorial superrotation of the planet atmosphere, which induces an angle shift of the planet reflection/emission phase modulation, as was suggested for Kepler-76 in the first paper of this series.  
We propose a modified BEER model that supports superrotation, assuming either a Lambertian or geometric reflection/emission phase function, and provides a photometry-consistent estimate of the planetary mass.
Our analysis shows that for Kepler-76 and HAT-P-7, the Lambertian superrotation BEER model is highly preferable over an unshifted null model, while for KOI-13 it is preferable only at a $1.4 \sigma$ level. For TrES-2 we do not find such preference.
For all four systems the Lambertian superrotation model mass estimates are in excellent agreement with the planetary masses derived from, or constrained by, radial velocity measurements. This makes the Lambertian superrotation BEER model a viable tool for estimating the masses of hot Jupiters from photometry alone.
We conclude that hot Jupiter superrotation may be a common phenomenon that can be detected in the visual  light curves of {\it Kepler}.
\end{abstract}


\section{Introduction}

The {\it Kepler} space telescope has produced more than $150,000$ nearly uninterrupted high-precision light curves \citep{koch10} that enable detection of minute astrophysical effects. As of 2014 June, analysis of these light curves yielded the discovery of more than 4200 planetary candidates \citep{kepexoplanets} through detection by the transit method \citep{batalha13}, of which more than 900 have been verified as planets by various methods \citep{exoplanetseu}. For such transiting planets, the orbital period, inclination, and radii of the star and planet, relative to the semimajor axis,  are directly measurable through analysis of the transit shape \citep[e.g.,][]{seager03}. However, there are additional astrophysical effects that produce flux variations along the orbital phase of a star-planet system, which depend on, and thus probe, additional properties of the planet. Such out-of-transit phase modulations are the result of three main stellar and planetary effects: BEaming, Ellipsoidal, and Reflection (BEER). 
The beaming effect, sometimes called Doppler boosting, causes an increase (decrease) of the brightness of any light source approaching (receding from) the observer \citep{rybicki79,loeb03}, with an amplitude proportional to the radial velocity (RV) of the source. Therefore, the stellar RV modulation due to a circular-orbit planet will produce a sine-like beaming phase modulation at the orbital period, if midtransit is defined as the phase zero point.
The ellipsoidal effect \citep{kopal59,morris85} is due to the tidal distortion of the star by the gravity of the planet \citep[e.g.,][]{loeb03,zucker07,mazeh08}, resulting in a cosine-like phase modulation at half the orbital period, for a circular-orbit planet under the same phase-zero definition. The amplitudes of the beaming and the ellipsoidal modulations for a transiting planet are {\it both} proportional to the planet mass, which cannot be probed by the transit method, thus providing an important insight into the planet composition.
The reflection/emission variation, on the other hand, is a result of light scattered off the planet dayside combined with light absorbed and later thermally reemitted by the planet atmosphere at different wavelengths \citep{vaz85,wilson90,maxted02,harrison03,for10,reed10}. This effect probes properties associated with the planet atmosphere response to its host-star radiation, such as the Bond albedo, scattered light geometric albedo, and heat redistribution parameters, among others. The reflection/emission phase modulation is expected to behave approximately as a cosine wave at the orbital period for a circular orbit.

In case the beaming, reflection, and ellipsoidal effects modulate as sine and cosine at the orbital period, and cosine at half the orbital period, respectively, their functions are orthogonal along the orbital phase, thus enabling the measuring of each of the effects amplitudes without interference from the other effects. As a result, the mass of a transiting planet can be independently estimated by either the beaming or the ellipsoidal amplitudes.
Such a derivation was performed for KOI-13 \citep{shporer11,mazeh12,esteves13}, HAT-P-7 \citep{esteves13}, TrES-2 \citep{barclay12,esteves13}, and Kepler-76b \citep{faigler13}. Interestingly, in all cases, except for the \cite{esteves13} analysis of TrES-2, the beaming-derived planetary mass estimate was significantly higher than the ellipsoidal-derived estimate. 
 In addition, RV measurements, available for HAT-P-7, TrES-2, and Kepler-76 \citep{winn09,odonovan06,faigler13}, show  spectroscopic RV amplitudes that are significantly smaller than the beaming-derived RV ones, pointing to puzzling inflated beaming amplitudes.

\citet{faigler13} suggested that  the inflated photometric beaming amplitude of Kepler-76 may be the result of 
a phase shift of the reflection signal, due to the superrotation phenomenon.
\citet{showman02} predicted, through three-dimensional atmospheric circulation model and simulations, that tidally locked, short-period planets develop fast eastward, or {\it superrotating}, equatorial jet streams that in some cases displace the hottest regions by $10^\circ{-}60^\circ$ longitude from the substellar point, resulting in a phase shift of the thermal emission phase curve of the planet. 
The existence of such a phase shift, due to superrotating equatorial jets, was confirmed through infrared phase curve observations of HD 189733 \citep{knutson07,knutson09,knutson12} that showed that the maximum flux occurred several hours before secondary eclipse.
Later, this phenomenon was further demonstrated by many numerical simulations \citep[e.g.,][]{showman08,showman09,thrastarson10,dobbs10,leconte13}.
In recent years significant progress has been made in understanding the superrotation phenomenon through semianalytic and linear approximation models \citep[e.g.,][]{gu09,gu11,watkins10,showman11,tsai14}.
Alternatively, high-altitude optically reflective clouds located westward of the substellar point may result in opposite-direction phase shift, as detected and explained by \citet{demory13} for Kepler-7b and also by \citet{angerhausen14} for Kepler-12b and Kepler-43b.
Close to the submission date of this paper \citet{esteves14} published a comprehensive phase curve analysis of $14$ {\it Kepler} hot Jupiters and found that in $7$ of them a phase shift of the planetary light offsets its peak  from the substellar point. 
 They concluded that eastward phase shifts dominate light from hotter planets (Kepler-76b and HAT-P-7b), while westward phase shifts dominate light from cooler planets (Kepler-7b, Kepler-8b, Kepler-12b, Kepler-41b and Kepler-43b).

The present paper extends the superrotation hypothesis by \citet{faigler13} and suggests that in addition to Kepler-76 this idea may be applicable to KOI-13, HAT-P-7, and TrES-2. We show that if such a superrotation-induced phase shift is present in the {\it Kepler} light curve, it should show up in the basic BEER phase curve model mainly as an apparently inflated beaming amplitude. We present the details and results of the new superrotation BEER model that provides a photometry-consistent estimate of the planetary mass.

This paper is organized as follows.
Section~2 presents the basic BEER model of a transiting planet assuming either a geometric or Lambertian reflection/emission phase function,
and Section~3 presents the superrotation BEER model, which models also the superrotation-induced phase shift of the reflection/emission modulation.
Section~4 then describes the analysis of the {\it Kepler} light curves of KOI-13, HAT-P-7, TrES-2, and Kepler-76;
Section~5 lists the parameters of the systems from the literature used in this paper and describes how additional stellar and planetary parameters were derived from them;
and Section~6 presents the results of the superrotation BEER models for the four systems.
Section~7 follows by discussing the relation between the {\it Kepler}-band-derived phase shift and the thermal emission phase shift,
Section~8 compares our results with those of previous studies
and Section~9 summarizes and discusses the findings of this work.

\section{The basic BEER model of a transiting planet}

We start with modeling the phase modulation of a circular-orbit transiting planet.
For such a planet we define the BEER model as a modification to the method described by \citet{faigler11}. First we define the orbital phase as
\begin{equation}
\phi=\frac{2 \pi}{P_{\rm orb}}\left( t-T_{0}\right)  ,
\end{equation}
where $P_{\rm orb}$ is the orbital period and $T_{0}$ is the midtransit time.
We then calculate, using robust linear fit \citep{holland77}, the first five Fourier series coefficients of the cleaned and detrended light curve \citep{mazeh10}, 
\begin{equation} \label{eq:fit}
\mathcal{M}(\phi)=a_0
-a_{1c}\cos \phi 
+a_{1s}\sin \phi 
-a_{2c} \cos 2\phi
-a_{2s}\sin 2\phi  ;
\end{equation}
where the signs are defined so that the coefficients are expected to be positive, though the fit can result in any sign for them.

In our approximation we express the relative flux modulation of the system due to a circular-orbit planet, as a result of the BEER effects, as
\begin{equation} \label{eq:effects}
\frac{\Delta F}{F}=a_0
+{A_{\rm ref}}\frac{\Phi\left( z \right)}{\sin i}
+A_{\rm beam}\sin \phi 
-A_{\rm ellip} \cos 2\phi  ,
\end{equation}
where $a_0$ is the relative flux zero point; $i$ is the orbital inclination angle; $z$ is the 
star--planet--observer angle; $\Phi\left(z \right)$ is the reflection/thermal emission phase function, which includes a $\sin i$ dependence; and $A_{\rm ref}$, $A_{\rm beam}$ and $A_{\rm ellip}$ are the reflection/emission, beaming and ellipsoidal semiamplitudes, respectively, which are expected to be positive. For the BEER effect amplitudes we use \citep{faigler11,zucker07,loeb03,morris93}
\begin{equation} \label{eq:amps}
A_\mathrm{ref} =
\alpha_\mathrm{refl}
\left(\frac{R_{\rm p}}{a} \right)^2 \sin i  ,
\end{equation}
\begin{equation*} 
A_\mathrm{beam} =
\alpha_\mathrm{beam}\, 4\frac{K_{\scriptscriptstyle\mathrm RV}}{c} 
=
2.7 \ \alpha_\mathrm{beam}
\left(\frac{M_*}{M_{\odot}}\right)^{-2/3}
\left(\frac{M_{\rm p}}{M_\mathrm{Jup}}\right) 
\left(\frac{P_\mathrm{orb}}{1 \,{\rm day}}\right) ^{-1/3}
\sin i \ {\rm (ppm)}  ,
\end{equation*}
\begin{equation*} 
 A_\mathrm{ellip}  =
\alpha_\mathrm{ellip}
\frac{M_{\rm p}}{M_*} \left(\frac{R_*}{a}\right)^3 \sin^2 i  ,
\end{equation*} 
where $K_{\scriptscriptstyle\mathrm RV}$ is the star RV semiamplitude; $a$ is the orbital semimajor axis; ${M_*}$, ${R_*}$, ${M_{\rm p}}$, and ${R_{\rm p}}$ are the mass and radius of the star and planet, respectively; and $\alpha_{\mathrm refl}$, $\alpha_\mathrm{beam}$, and $\alpha_\mathrm{ellip}$ are the reflection/emission, beaming, and ellipsoidal coefficients, respectively. It is important to note that $\alpha_{\mathrm refl}$ encapsulates two distinct and different planet luminosity sources. One is due to the planet dayside geometric albedo resulting in reflected-light phase modulation, and the other is due to the planet day--night temperature contrast, resulting in thermal-emission phase modulation. While we expect both modulations to be proportional to $\left(\frac{R_{\rm p}}{a} \right)^2$, it is only a simplifying assumption to use for both the same phase function $\Phi\left(z \right)$.

The reflection/emission phase function depends on the $z$ angle, defined as the 
star--planet--observer angle,
which is related to the $\phi$ phase through
\begin{equation} \label{eq:z}
\cos z=-\sin i\cos \phi \  \Rightarrow \ \cos 2z=\sin^2 i\cos 2\phi +  {\rm constant \ term}  ,
\end{equation}
where throughout this discussion we ignore constant terms that are not phase dependent, as these add up to the total flux and are not measurable from the data.

A possible choice for the phase function is the geometric reflection function, which assumes that the received flux is proportional to the projected area on the sky plane of the illuminated half-sphere of the planet, as seen by the observer. Following a notation similar to \citet{mislis12}, the geometric reflection phase function is
\begin{equation} \label{eq:geodef}
\frac{\Phi_{\rm geo}(z)}{\sin i}= \frac{\cos z}{\sin i} = -\cos \phi  .
\end{equation}
 Under this definition of the phase function, the BEER  amplitudes (Equation~(\ref{eq:effects})) are directly related to the Fourier coefficients measured from the light curve (Equation~(\ref{eq:fit})) through 
\begin{equation} \label{eq:geo}
\{ A_{\rm ref}=a_{1c} \ ,\  A_{\rm beam}=a_{1s} \ ,\ A_{\rm ellip,geo}=a_{2c} \ ,\  a_{2s}=0 \} \ .
\end{equation}

It is more common, however, to model the planet as a Lambert sphere \citep{lambert60,russell16,sobolev75,demory11}, 
which assumes that the planet surface is an ideal diffuse reflector, i.e., of equal reflection to all directions in the half-sphere facing the star, regardless of the incident light direction. The resulting Lambertian reflection phase function is
\begin{equation}
\Phi_{\rm Lam} =  \frac{2}{\pi}\left(\sin |z| +(\pi-|z|)\cos |z| \right) \ , \ \ 
\{-\pi \le z \le \pi\}  ,
\end{equation}
where we have defined $\Phi_{\rm geo}$ and $\Phi_{\rm Lam}$ with the same peak-to-peak amplitude.
Evaluating the Fourier series expansion of $\Phi_{\rm Lam}$, we realize that for all integers $n$ the $\sin nz$ coefficients equal zero, as this function is symmetric about the $z=0$ point. 
Therefore, expanding with the cosine functions, we get 
\begin{equation}
\Phi_{\rm Lam} =  \frac{8}{\pi^2}+\cos z +\frac{16}{9\pi^2}\cos 2z +\frac{16}{225\pi^2}\cos 4z + \rm smaller \ terms \ .
\end{equation}
Ignoring all harmonics higher than $\cos 2z$
provides accuracy better than $1\%$, which gives, after translating from the $z$ angle to the $\phi$ angle,
\begin{equation} \label{eq:LambApp}
\frac{\Phi_{\rm Lam}}{\sin i} \approxeq -\cos \phi +0.18 \sin i \cos 2\phi 
+ \rm a \ constant \ term \ .
\end{equation}
The resulting $\Phi_{\rm Lam}$ form shows that geometric reflection is simply a first harmonic approximation of Lambertian reflection and  that Lambertian reflection has a cosine component in the {\it second} harmonic.
Next, from Equations~(\ref{eq:effects}) and (\ref{eq:LambApp}) we get
\begin{equation} \label{eq:BEER}
\frac{\Delta F_{\rm Lam}}{F}=a_0
- A_{\rm ref}\cos \phi 
+A_{\rm beam}\sin \phi 
-\left(A_{\rm ellip,Lam}-0.18 A_{\rm ref} \sin i\right) \cos 2\phi  ,
\end{equation}
which enables deriving the relations between the BEER amplitudes and the measured Fourier coefficients, resulting in 
\begin{equation} \label{eq:lamb}
\{A_{\rm ref}=a_{1c} \ ,\  A_{\rm beam}=a_{1s} \ ,\  A_{\rm ellip,Lam}=a_{2c}+0.18 a_{1c}\sin i \ ,\  a_{2s}=0 \}\ .
\end{equation}
We see that the apparent unnatural definition of the reflection term in Equation~(\ref{eq:effects}) actually leads to a simple form for the BEER effect amplitudes, resulting in the same $A_{\rm ref}$ value for geometric and Lambertian reflection, representing half the peak-to-peak variation of the reflection effect in both cases. On the other hand, as demonstrated by \citet{mislis12}, the Lambertian reflection assumption results in a larger ellipsoidal semiamplitude $A_{\rm ellip,Lam}$, relative to $A_{\rm ellip,geo}$ in the geometric case. In this paper we consider the two alternative ellipsoidal semiamplitudes using Equations~(\ref{eq:geo}) and (\ref{eq:lamb}).


\section{The superrotation BEER model of a transiting planet}   \label{sec:SR}      %
To model the superrotation-induced phase shift, we follow the model suggested by \citet{faigler13}, while extending it to either geometric or Lambertian phase functions.
To account for superrotation in our analysis, we adopt a simplistic model for the total reflection/emission modulation that is the sum of a scattered-light phase function and a {\it phase-shifted} emission phase function. In this model we represent the total reflection/emission modulation in the {\it Kepler} band as a phase-shifted Lambertian or geometric phase function. While accurate for geometric scattered-light and emission phase functions, it is only an approximation for Lambertian phase functions. Under these model assumptions, we simply need to replace $\phi$ with $\phi+\delta_{\rm sr}$ in the geometric or Lambertian phase function   (Equation~(\ref{eq:geodef}) or (\ref{eq:LambApp})), 
where $\delta_{\rm sr}$ is the phase shift in the {\it Kepler} band due to superrotation, assumed to be positive. 
Inserting each shifted phase function into Equation~(\ref{eq:effects}), we have for geometric reflection
\begin{equation} \label{eq:geoSR}
\frac{\Delta F_{\rm C,SR}}{F}=a_0
- A_{\rm ref}\cos\delta_{\rm sr}\cos \phi 
+\left(A_{\rm beam}+A_{\rm ref}\sin\delta_{\rm sr}\right)\sin \phi 
-A_{\rm ellip}\cos 2\phi \ ,
\end{equation}
\begin{equation*} \label{eq:geoSRa}
\Longrightarrow \ \{ A_{\rm ref}\cos\delta_{\rm sr}=a_{1c} \ ,\  A_{\rm beam}+\underline{A_{\rm ref}\sin\delta_{\rm sr}}=a_{1s} \ ,\ A_{\rm ellip}=a_{2c} \ ,\  a_{2s}=0 \} \ ,
\end{equation*}
or for Lambertian reflection
\begin{equation} \label{eq:LambSR}
\frac{\Delta F_{\rm L,SR}}{F}=a_0
- A_{\rm ref}\cos\delta_{\rm sr}\cos \phi 
+\left(A_{\rm beam}+A_{\rm ref}\sin\delta_{\rm sr}\right)\sin \phi 
\end{equation}
\begin{equation*}
-\left(A_{\rm ellip}-0.18 A_{\rm ref}\cos2\delta_{\rm sr} \sin i\right) \cos 2\phi
-0.18 A_{\rm ref}\sin2\delta_{\rm sr} \sin i\ \sin 2\phi \ ,
\end{equation*}
\begin{equation*} \label{eq:LambSRa}
\Longrightarrow \ \{ A_{\rm ref}\cos\delta_{sr}=a_{1c} \ ,\  A_{\rm beam}+\underline{A_{\rm ref}\sin\delta_{\rm sr}}=a_{1s} \ ,
\end{equation*}
\begin{equation*}
A_{\rm ellip}-0.18 A_{\rm ref}\cos2\delta_{\rm sr} \sin i=a_{2c} \ ,\  0.18 A_{\rm ref}\sin2\delta_{\rm sr} \sin i =a_{2s} \} \ .
\end{equation*}
We see that for both phase functions, a phase shift in the reflection modulation results in the additional underlined term of ${A_{\rm ref}\sin\delta_{\rm sr}}$, which inflates the $\sin \phi$ coefficient and {\it might be wrongly interpreted as an inflated beaming amplitude}. In addition, assuming Lambertian reflection yields additional smaller corrections to the $\cos 2\phi$ coefficient and to the previously assumed-to-be-zero $\sin 2\phi$ coefficient (see Equation~(\ref{eq:BEER}) versus Equation~(\ref{eq:LambSR}) ). 
As a summary, Table~\ref{table:amps} lists, for the different BEER model types, the relations between the astrophysical effect amplitudes $\{A_{\rm beam},A_{\rm ellip},A_{\rm ref}\}$ and the measured Fourier coefficients $\{a_{1c},a_{1s},a_{2c},a_{2s}\}$. 

\clearpage
 
\begin{table}[!h]
\caption{ Relations between BEER models amplitudes and Fourier coefficients}
\hfill{ }\\
\small \begin{tabular}{|l|c|c|c|c|}
\hline
Fourier coefficients $\rightarrow$ & $\cos \phi$ &  $\sin \phi$ &  $\cos 2\phi$ &  $\sin 2\phi$ \\
BEER models $\downarrow$      & $(-a_{1c})$ &  $(a_{1s})$ &  $(-a_{2c})$ &  $(-a_{2s})$ \\

\hline
Geometric  & $-A_{\rm ref}$ & $A_{\rm beam}$ & $-A_{\rm ellip}$ & $0$ \\
Reflection & & & & \\
\tableline
Lambert & $-A_{\rm ref}$ & $A_{\rm beam}$ & $-A_{\rm ellip}$ & $0$ \\
Reflection & & & $+0.18 A_{\rm ref}\sin i$ & \\
\hline
\hline
Geometric  & $-A_{\rm ref}\cos \delta_{\rm sr}$ & $A_{\rm beam}+A_{\rm ref}\sin \delta_{\rm sr}$ & $-A_{\rm ellip}$ & $0$ \\
Superrotation & & & & \\
\hline
Lambert  & $-A_{\rm ref}\cos \delta_{\rm sr}$ & $A_{\rm beam}+A_{\rm ref}\sin \delta_{\rm sr}$ & $-A_{\rm ellip}$ & $-0.18A_{\rm ref}\sin 2\delta_{\rm sr}\sin i$ \\
Superrotation & & & $+0.18A_{\rm ref}\cos 2\delta_{\rm sr}\sin i$ & \\
\hline
\end{tabular}
\label{table:amps}
\end{table}


\section{Photometric analysis} \label{sec:photo}
In this section we describe the analysis of the {\it Kepler} long-cadence Pre-search Data Conditioning (PDC) light curves of the Q2 to Q16 quarters, spanning $1302$ days, for KOI-13, HAT-P-7, TrES-2, and Kepler-76. The data were first cleaned and detrended following the methods described by \citet{mazeh10} and \citet{faigler13}. We then fitted the data using Equation~(\ref{eq:fit}) and derived the Fourier coefficients, while masking out data points in or around the transits and occultations. To test the robustness of our process, we performed the same analysis on the raw {\it Kepler} light curves of the four systems, yielding no significant differences between the results of the two analyses.

We have paid special attention in the fitting process to deriving realistic uncertainties for the Fourier coefficients. To do that, we performed the fitting for each {\it Kepler} quarter separately, and we report the best-fit coefficient as  $a=median\{a_q\}$, where $\{a_q\}$ are the fit results over the {\it Kepler} quarters. Next, we estimated the uncertainty from the scatter of $\{a_q\}$, using a modification to the Median Absolute Deviation method, as 
$\sigma_a=1.253 \times {\rm mean\{|{a_q-a}|\}}/{\sqrt{N_q}}$,
 where $N_q$ is the number of {\it Kepler} quarters for which data are available. This calculation should result in uncertainties similar to linear fitting for uncorrelated Gaussian noise, while providing more realistic uncertainties for correlated noise or systematic effects.
Considering the quarter-to-quarter variation is supported by \citet{vaneylen13}, who 
measured seasonal variations of about $1\%$ of the transit depth of HAT-P-7 over the {\it Kepler} quarters.
Even more relevant to our case of periodic modulations, they also measured about $1\%$ seasonal variations of the pulsation amplitude of the RR Lyr star KIC 6936115. In both cases \citet{vaneylen13} showed that the seasonal variations were over an order of magnitude larger than the naive uncertainties derived from fitting the combined {\it Kepler} light curve of all available quarters.
Although we do not see any correlation in amplitudes measured in same season quarters (i.e. separated by $1$ yr), we do measure quarter-to-quarter variations that are significantly larger than the fitting process uncertainties.
 Indeed, our reported uncertainties, which are derived from the quarter-to-quarter variations, are usually larger than those reported by other authors for the same quantities, but we believe that they better capture the uncertainty embedded in the data.

For KOI-13 we inflated the amplitudes by a third-light factor of 1.82 that was estimated by \citet{szabo11}, while for the other systems we used the Kepler Input Catalog (KIC) third-light estimates. The KIC third-light average estimates for HAT-P-7, TrES-2, and Kepler-76 are $0.2\%$, $0.8\%$, and $5.7\%$, respectively, and incorporating or ignoring them had negligible effect on our results.  

The fitted Fourier coefficients of the first two orbital-period harmonics, after correction for third light, are listed in Table~\ref{table:photo}. 

\begin{table}[!h]
\caption{Derived Fourier coefficients}
\hfill{ }\\
\begin{tabular}{|l|c|c|c|c|c|}
\tableline
Fourier coefficients $\rightarrow$  & $\cos \phi$ &  $\sin \phi$ &  $\cos 2\phi$ &  $\sin 2\phi$ & \\
System $\downarrow$                  & $(-a_{1c})$ &  $(a_{1s})$ &  $(-a_{2c})$ &  $(-a_{2s})$ & \\
\tableline
KOI-13 &$-71.0 \pm 0.7$ & $8.2 \pm 0.7$ & $-55.9 \pm 0.8$ & $-2.0 \pm 1.1$ & ppm\\
\tableline
HAT-P-7 & $-32.2 \pm 0.9$ & $6.6 \pm 1.1$ & $-14.8 \pm 1.2$ & $-0.4 \pm 0.6$ & ppm\\
\tableline
TrES-2   & $-1.5 \pm 1.1$ & $1.9 \pm 1.3$ & $-2.9 \pm 0.6$ & $-0.6 \pm 0.6$ & ppm\\
\tableline
Kepler-76 & $-54.4 \pm 2.3$ & $13.1 \pm 1.3$ & $-12.7 \pm 1.6$ & $-2.8 \pm 0.8$ & ppm\\
\tableline
\end{tabular}
\label{table:photo}
\end{table}


\section{Systems parameters from literature}
For transiting planets the orbital period $P_{\rm orb}$, inclination angle $i$, and the ratio of primary radius to orbital semimajor axis $R_*/a$ are directly measurable from the transit light curve. When combined with a stellar model for the primary mass $M_*$ and the effect coefficients $\alpha_{\rm beam}$ and $\alpha_{\rm ellip}$, which also depend on the stellar parameters, the planetary mass can be independently estimated from either the beaming or the ellipsoidal amplitude, using Equation~(\ref{eq:amps}).
To estimate the planetary mass from the different models, we used the systems parameters available in the literature.
The upper section of Table~\ref{table:sys} lists the parameter values used from the literature for the four systems. 
The lower section of the table lists additional parameters that we derived from the literature parameters listed in the upper section. 

We estimated $\alpha_{\rm beam}$ by numerically shifting spectra from the library of \citet{castelli04} models against the {\it Kepler} response function following \citet{faigler12}, while taking into account the photon-counting nature of {\it Kepler} \citep{bloemen11}. The $\alpha_{\rm ellip}$ coefficient was estimated using the interpolated limb and gravity darkening coefficients from \citet{claret11} and the stellar parameters, with the \citet{morris93} equation \citep[see][]{mazeh10}. 

To estimate the maximum fraction of the reflection/emission amplitude originating from thermal reemission, we follow \citet{cowan11} and derive the no-albedo, no-redistribution, effective dayside temperature $T_{\epsilon=0}$, which translates in the {\it Kepler} band to the maximum emission amplitude $A_{\rm ref,\epsilon=0}$, both listed in Table~\ref{table:sys} for the four systems.

KOI-13 is a hierarchical triple stellar system, where KOI-13A and KOI-13B are a common proper-motion fast-rotating A-type stars ($V_{\rm A}=9.9$, $V_{\rm B}=10.2$) with $\sim$$1\arcsec.2$ angular separation, and KOI-13C is a $0.4$--$1 M_{\odot}$ star on a $65.8$ day orbit around KOI-13B \citep{aitken04,dommanget94,szabo11,santerne12}.
KOI-13b is a $\sim$$1.4 R_{\rm Jup}$ planet on a $1.76$ day orbit around the system main component KOI-13A \citep{szabo11,barnes11,santerne12,batalha13,shporer14}.
\citet{santerne12} determined, through spectroscopic RV observations of the system, a $3 \sigma$ upper limit of $14.8 M_{\rm Jup}$ for the mass of KOI-13b.
 For this system we used the transit derived parameters from \citet{barnes11}, which successfully modeled the asymmetry of the KOI-13 transit light curve assuming a gravity-darkened rapidly rotating host star in order to constrain the system's spin-orbit alignment and transit parameters.  

HAT-P-7b is a $1.8 M_{\rm Jup}$, $1.5 R_{\rm Jup}$ planet on a $2.2$ day retrograde orbit around a $9.7$ mag evolved F6 star \citep{pal08,winn09}.
For this system we used the transit derived parameters from \citet{welsh10} and the stellar parameters derived through asteroseismology by \citet{christensen10}.
 
TrES-2b is the first transiting planet discovered in the {\it Kepler} field \citep{odonovan06}. It is a $1.17 M_{\rm Jup}$, $1.16 R_{\rm Jup}$ planet on a $2.47$ day orbit around a $11.3$ mag G0V star.
For this system we used the transit-derived parameters and the asteroseismology-derived stellar parameters from \citet{barclay12}.

Kepler-76b is a $2 M_{\rm Jup}$, $1.25 R_{\rm Jup}$ planet orbiting a $13.3$ mag F star in $1.55$ days.
For this system we used the transit-derived parameters and spectroscopic stellar parameters from \citet{faigler13}.

%
\capstartfalse
\begin{deluxetable}{lrrrrl}
\tabletypesize{\scriptsize}
\tablecaption{Systems parameters from literature \label{table:sys}}

\tablewidth{0pt}
\tablehead{
& \colhead{KOI-13}  & \colhead{HAT-P-7} &  \colhead{TRES-2} &\colhead{Kepler-76}
}

\startdata
$T_{\rm eff}$  (K) & $8500 \pm 400$$^{\rm a}$   & $6350 \pm 80$$^{\rm e}$  & $5850 \pm 50$$^{\rm i}$ & $6300 \pm 200$$^{\rm l}$ & Host star effective temperature               \\
$M_*$ ($ M_{\odot}$) & $2.05 \pm 0.2$$^{\rm a,b} $  & $1.53 \pm 0.04$$^{\rm f}$ & $0.94 \pm 0.05$$^{\rm j}$  & $1.2 \pm 0.2$$^{\rm l}$  & Host star mass           \\
$\rm [m/H]$ (dex) & $0.2 \pm 0.1$$^{\rm a,b} $  & $0.3 \pm 0.1$$^{\rm e}$ & $-0.1 \pm 0.1$$^{\rm i}$  & $-0.1 \pm 0.2$$^{\rm l}$  & Host star metallicity          \\
$R_*/a$  & $0.2237 \pm 0.0041$$^{\rm m}$  & $0.241 \pm 0.001$$^{\rm h}$ & $0.126 \pm 0.001$$^{\rm j}$  & $0.2209 \pm 0.0027$$^{\rm l}$ & Fractional primary radius             \\
$R_{\rm p}/a$  & $0.0189 \pm 0.0004$$^{\rm m}$  & $0.0187 \pm 0.0001$$^{\rm h}$ & $0.0158 \pm 0.0001$$^{\rm j}$  & $0.0214 \pm 0.0008$$^{\rm l}$ & Fractional planet radius             \\
Inclination (deg)& $85.9$$^{\rm m} $ & $83.1$$^{\rm h}$ & $83.9$$^{\rm j}$ & $78.0$$^{\rm l}$ & Orbital inclination \\
Period (days)    & $1.7635877 $$^{\rm c}$& $2.20473$$^{\rm h}$ & $2.47061320$$^{\rm j}$ & $1.54492875$$^{\rm l}$  & Orbital period \\ 
$K_{\scriptscriptstyle \mathrm RV} ({\rm m/s})$ &$<1.3$$^{\rm d}$ &$212 \pm 5$$^{\rm g}$ &$181.3 \pm 2.6$$^{\rm k}$ & $306 \pm 20$$^{\rm l}$ & Spectroscopic RV semiamplitude \\
\tableline
$M_{\rm p,RV}$ ($ M_{\rm Jup}$) & $<14.8$$^{\rm d}$   & $1.82 \pm 0.05$   & $1.17 \pm 0.04$ & $2.0 \pm 0.26$   & Planet mass derived from RV          \\

$\alpha_{\rm beam}$ & $0.63 \pm 0.05$ & $0.91 \pm 0.04$ &  $0.99 \pm 0.04$ & $0.92 \pm 0.04$ &  Beaming coefficient \\
$\alpha_{\rm ellip}$ & $1.53 \pm 0.08$& $1.21 \pm 0.03$ & $1.31 \pm 0.03$ & $1.22 \pm 0.03$  &  Ellipsoidal coefficient \\

$T_{\epsilon=0} $ (K) & 3630 &2800 & 1880 & 2670 & Planet max. dayside temperature \\
$A_{\rm ref,\epsilon=0} $ (ppm) & 82 & 34 & 3.4 & 37 & Planet max. emission semiamplitude\\

\tableline

\enddata
\tablenotetext{\space}{ $^{\rm a}$\citet{szabo11}.} 
\tablenotetext{\space}{$^{\rm b}$Uncertainties added by authors.}
\tablenotetext{\space}{$^{\rm c}$\citet{batalha13}.  }
\tablenotetext{\space}{ $^{\rm d}$$3 \sigma$ upper limit \citep{santerne12}.}
\tablenotetext{\space}{ $^{\rm e}$\citet{pal08}.}
\tablenotetext{\space}{$^{\rm f}$\citet{christensen10}.}
\tablenotetext{\space}{$^{\rm g}$\citet{winn09}.}
\tablenotetext{\space}{$^{\rm h}$\citet{welsh10}.} 
\tablenotetext{\space}{ $^{\rm i}$\citet{sozzetti07}.}
\tablenotetext{\space}{$^{\rm j}$\citet{barclay12}.}
\tablenotetext{\space}{$^{\rm k}$\citet{odonovan06}.}
\tablenotetext{\space}{$^{\rm l}$\citet{faigler13}. }
\tablenotetext{\space}{$^{\rm m}$\citet{barnes11}.}
\end{deluxetable}
\capstarttrue
\clearpage

\section{Results}
We are now in a position to estimate the planetary mass using the different models and compare it with the mass estimate derived from the RV semiamplitude $M_{\rm p,RV}$. For the unshifted geometric/Lambert reflection models, the beaming/ellipsoidal-based mass estimates can be derived directly from their respective semiamplitudes using Equation~(\ref{eq:amps}), Table~\ref{table:amps}, and Table~\ref{table:photo}.
For the geometric/Lambert superrotation models we fitted the light-curve Fourier coefficients, using the system parameters and the superrotation model equations, while deriving the planetary mass $M_{\rm p,sr}$, reflection coefficient $\alpha_{\mathrm refl}$, and phase shift $\delta_{\rm sr}$, which minimize the $\chi^2$ of the fit.
Table~\ref{table:SR} lists the planetary masses, phase-shift angles, and reflection coefficient estimates assuming the different models and their corresponding $\chi^2$ values. 
 For Kepler-76, HAT-P-7, and KOI-13, $\chi^2$-tests show that the Lambert superrotation BEER model is preferable over a zero-phase-shift null model, with confidence levels of $7.4\sigma$, $3.3\sigma$, and $1.4\sigma$, respectively. For TrES-2 the Lambert superrotation model is not preferable over the unshifted model, resulting in a phase shift that is consistent with zero.
For the three detections the resulting superrotation phase shift angle is small and well within the theoretical limit of $60^\circ$ predicted by \citet{showman02}. Also, for all four systems the mass estimate derived from the Lambert superrotation BEER model is consistent the RV-based planetary mass, indicating that the Lambert superrotation model resolves the inconsistency and provides a good photometric estimate for the planet mass, derived {\it solely} from the {\it Kepler} photometry, given a good stellar model.

The  $\chi^2$-tests are valid in these cases as we assume that our measured amplitude uncertainties are well estimated (see Section~\ref{sec:photo}). To verify this claim, we also fitted the same BEER models to the out-of-transit data points and calculated the {\it F}-test confidence levels of the fits. For the three detections KOI-13, HAT-P-7 and Kepler-76 the {\it F}-test confidence levels were better than the $\chi^2$-test confidence levels, both indicating preference for the superrotation models. 

Interestingly, Table~\ref{table:SR} shows that the planetary mass derived by the superrotation model $M_{\rm p,sr}$ is very close to the mass derived directly from the ellipsoidal amplitude $M_{\rm p,ellip}$. This is because introducing the additional phase-shift parameter $\delta_{\rm sr}$ into the model can significantly modify the beaming amplitude $A_{\rm beam}$ while keeping the ellipsoidal amplitude $A_{\rm ellip}$ unchanged for geometric phase function, or almost unchanged for Lambertian phase function (see Table~\ref{table:amps}). As a result the superrotation model best fit will converge to a phase-shift value that modifies the beaming amplitude so that its resulting planetary mass aligns with the ellipsoidal-derived mass.

\begin{figure*} 
\centering
\resizebox{17cm}{8.0cm}
{
\includegraphics{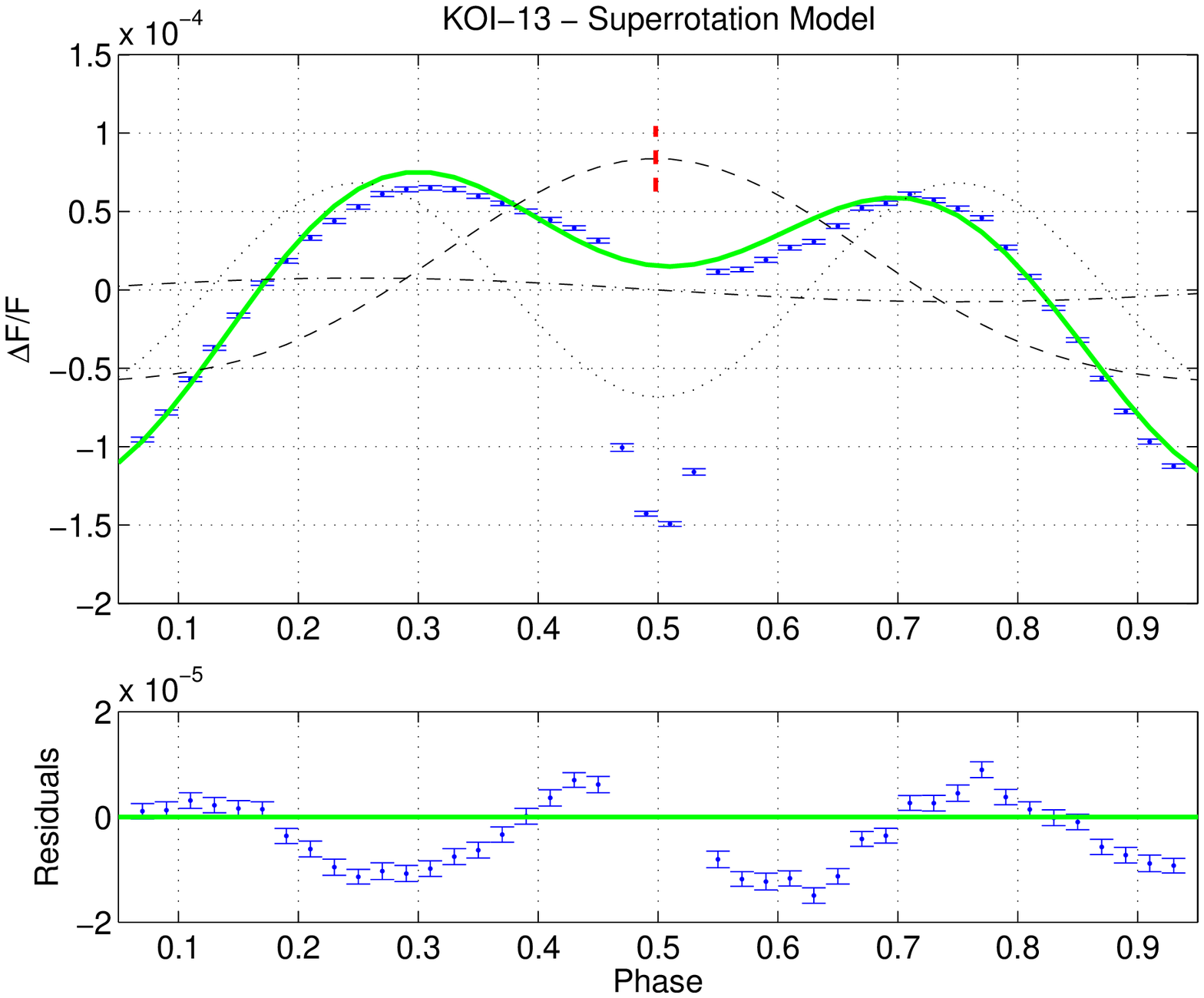}  
\includegraphics{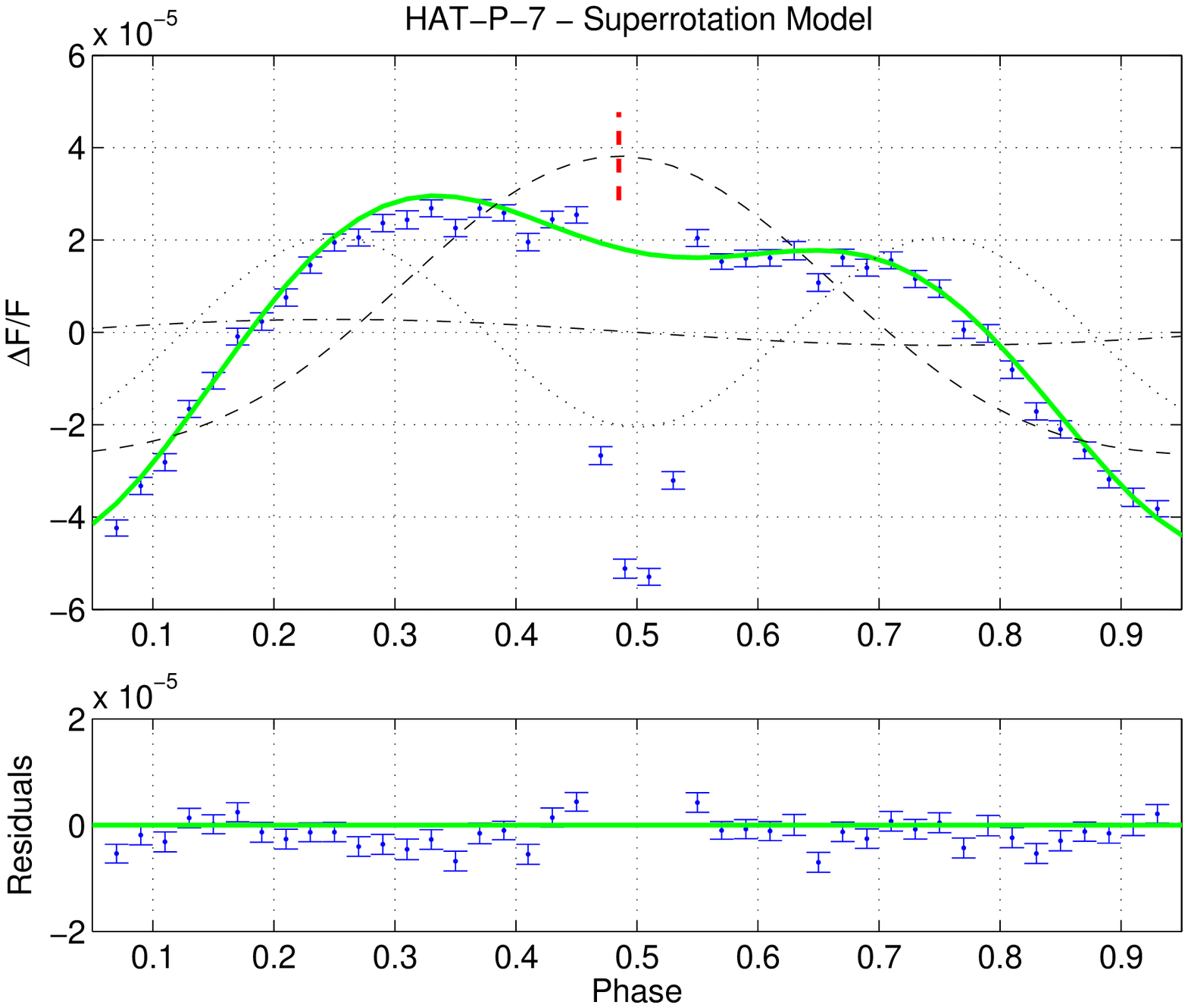}  
}
\centering
\resizebox{17cm}{8.0cm}
{
\includegraphics{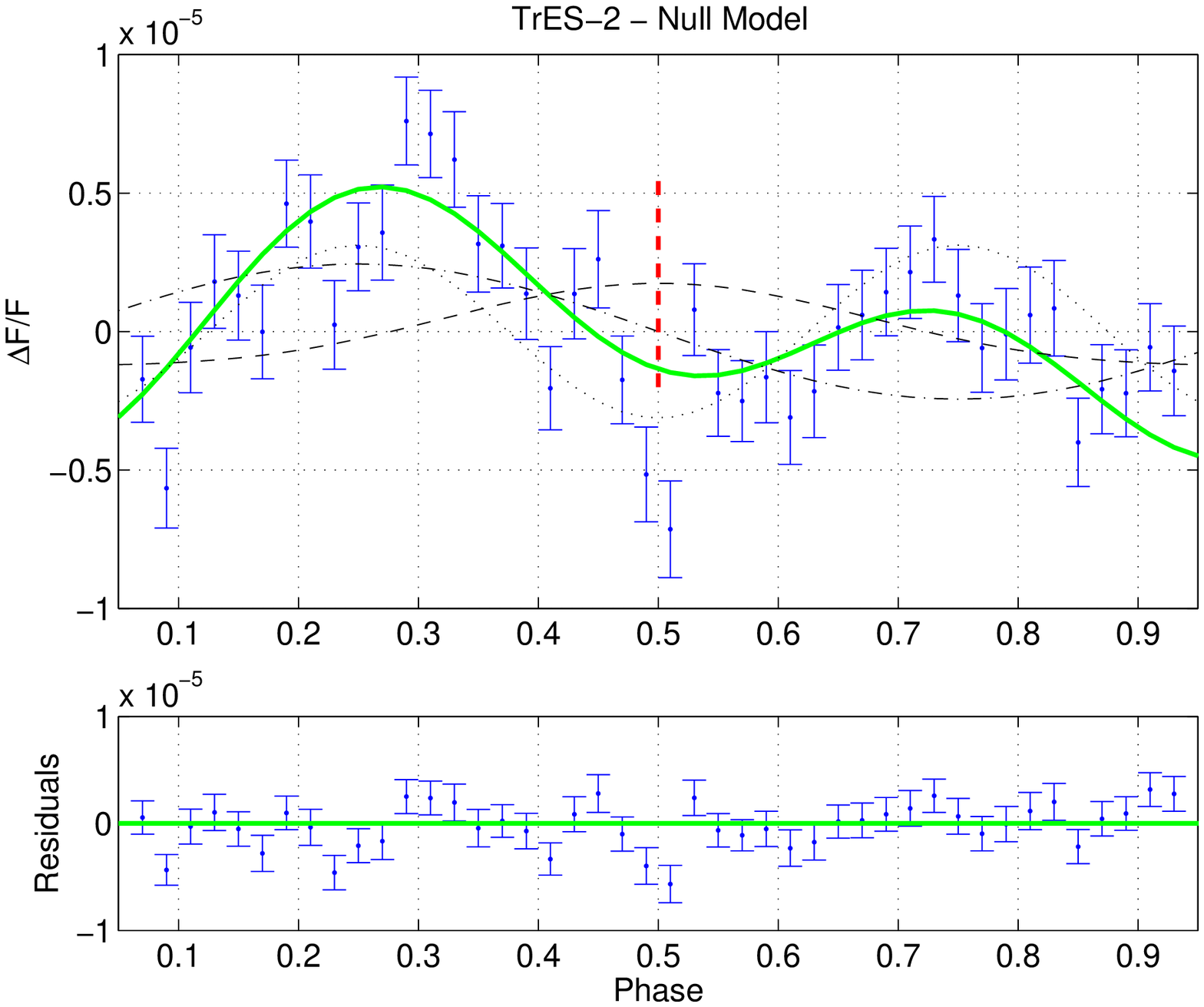}  
\includegraphics{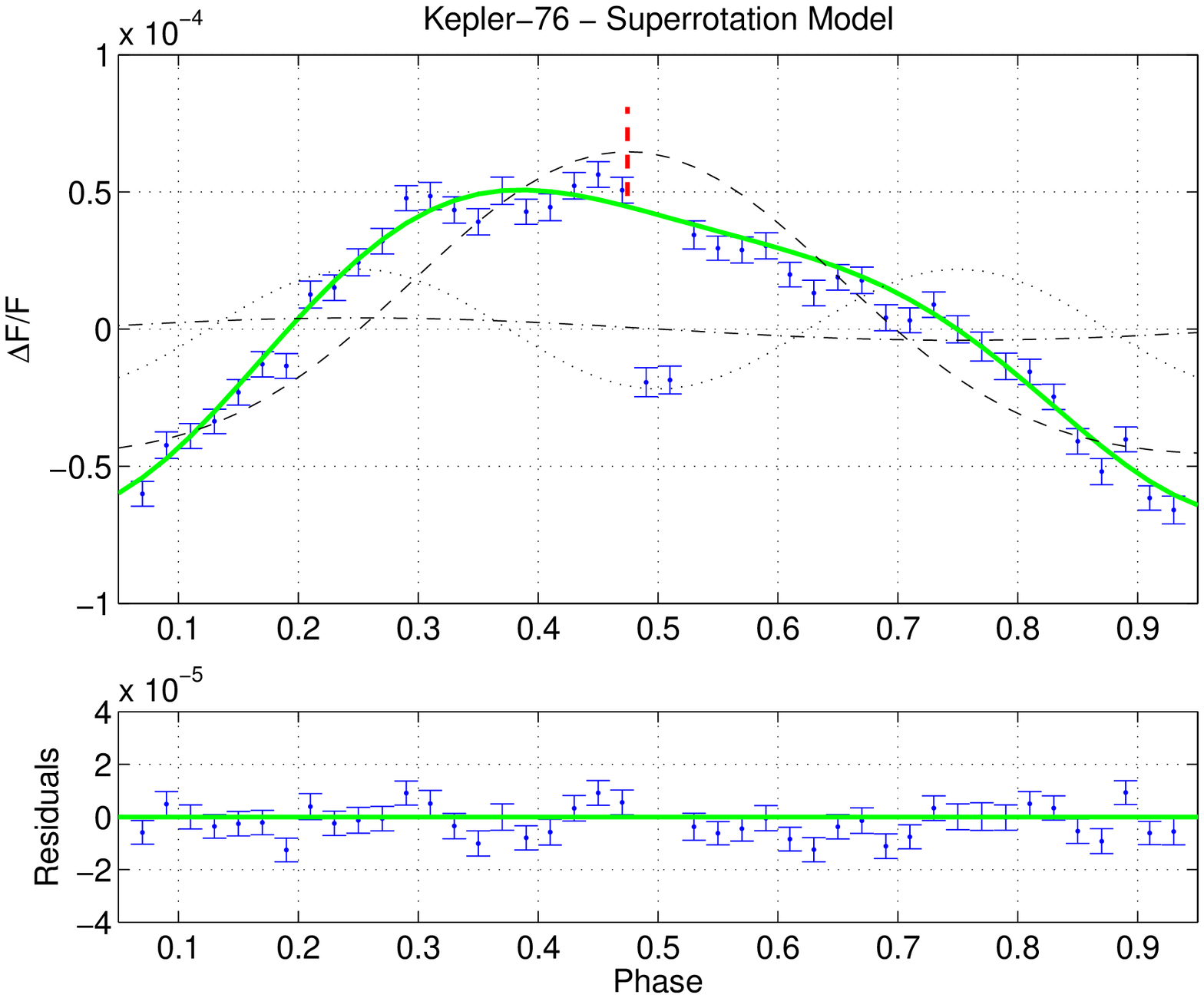}  
}
\caption{
Binned, cleaned, and detrended data points and best-fit preferred model of the four systems, folded at the orbital period. In the top panel of each plot, the solid line presents the Lambert superrotation model for Kepler-76, HAT-P-7, and KOI-13 and the unshifted BEER model for TrES-2, and the dots with error bars present the binned data points. The dashed, dot-dashed, and dotted lines present the Lambert reflection/emission, beaming, and ellipsoidal models, respectively. The vertical red dashed lines mark the phases of maximum reflection/emission, which are $0.474$, $0.485$, and $0.498$  for Kepler-76, HAT-P-7, and KOI-13, respectively.
The residuals are plotted in the bottom panel.  Note the different scales of the top and bottom panels of each plot. 
}
\label{figure:folds}
\end{figure*}

Figure~\ref{figure:folds} presents the cleaned and detrended data points, folded at the orbital period and grouped into 50 phase bins, and the best-fit preferred models of the four systems. The figure also shows the Lambert reflection/emission, beaming, and ellipsoidal models and marks the phase of the maximum reflection/emission modulation, which for Kepler-76, HAT-P-7, and KOI-13 is smaller than 0.5 owing to the superrotation phase shift. Note, however, that the model fitting was performed on the derived Fourier coefficients and not directly on the data points. The folded and binned light-curve data are plotted here for illustrating the periodic modulation.  

\capstartfalse
\begin{deluxetable}{lrrrrl}
\tabletypesize{\scriptsize}
\tablecaption{Planetary mass estimates and superrotation phase shift angle \label{table:SR}}

\tablewidth{0pt}
\tablehead{
& \colhead{KOI-13}  & \colhead{HAT-P-7} &  \colhead{TRES-2} &\colhead{Kepler-76}
}

\startdata

$M_{\rm p,RV}$ ($ M_{\rm Jup}$) & $<14.8$ $^{\rm f}$   & $1.82 \pm 0.05$   & $1.17 \pm 0.04$ & $2.0 \pm 0.26$   & Planet mass derived from RV          \\
\\
\tableline
\\

Geometric reflection: & & & & &  Unshifted geometric reflection\\
$M_{\rm p,beam}$ ($ M_{\rm Jup}$) & $9.4 \pm 1.3$   & $4.6 \pm 0.8$  & $0.90 \pm 0.66$   & $7.0 \pm 1.2$ & Planet mass from beaming     \\
$M_{\rm p,ellip}$ ($ M_{\rm Jup}$)& $7.0 \pm 0.9$   & $1.42 \pm 0.12$ & $1.11 \pm 0.23$   & $1.25 \pm 0.28$  & Planet mass from ellipsoidal          \\
$\chi^2_{\rm null}$ & 14.1 &17.5 & 7.7 & 74.1 &  $\chi^2$ of unshifted null model \\
\\
\tableline
\\

Lambert reflection:& & & & & Unshifted Lambert reflection\\
$M_{\rm p,beam}$ ($ M_{\rm Jup}$) & $9.4 \pm 1.3$   & $4.6 \pm 0.8$  & $0.90 \pm 0.66$   & $7.0 \pm 1.2$ & Planet mass from beaming     \\
$M_{\rm p,ellip}$ ($ M_{\rm Jup}$)& $8.6 \pm 1.1$   & $1.97 \pm 0.14$ & $1.22 \pm 0.25$  & $2.21 \pm 0.43$  & Planet mass from ellipsoidal          \\
$\chi^2_{\rm null}$ & 4.2 &12.2 & 1.4 & 55.5 &  $\chi^2$ of unshifted null model \\
\\
\tableline
\\

Geometric superrotation:& & & & & Shifted geometric reflection\\
$M_{\rm p,sr}$ ($ M_{\rm Jup}$) & $7.0 \pm 0.9$ & $1.42 \pm 0.13$ & $1.11 \pm 0.23$ & $1.25 \pm 0.28$   & Planetary mass        \\
$\delta_{\rm sr}$ (deg) & $1.7 \pm 0.8$   & $8.0 \pm 2.0$  & $-12 \pm 51$  & $11.2 \pm 1.5$      & phase shift angle       \\
$\alpha_{\rm refl}$  & $0.20 \pm 0.01$ & $0.093 \pm 0.003$ & $0.006 \pm 0.004$ & $0.12 \pm 0.01$      & Reflection coefficient       \\

$\chi^2$ &6.2 & 0.5 & 6.9 & 13.2 &  $\chi^2$ of the model \\
$P_{\rm}$ &$\rm4.7E-3$ $(2.8\sigma)$ & $\rm3.7E-5$ $(4.1\sigma)$ & $\rm3.8E-1$ $(0.9\sigma)$& $\rm 5.9E-15$ $(7.8\sigma)$ & $\chi^2$-test confidence level \\ 

\\
\tableline
\\

Lambert superrotation:& & & & & Shifted Lambert reflection\\
$M_{\rm p,sr}$ ($ M_{\rm Jup}$) & $8.6 \pm 1.1$ & $1.97 \pm 0.14$ & $1.13 \pm 0.24$ & $2.18 \pm 0.42$   & Planetary mass        \\
$\delta_{\rm sr}$ (deg) & $0.8 \pm 0.9$   & $5.4 \pm 1.5$  & $13 \pm 54$  & $9.2 \pm 1.3$      & phase shift angle       \\
$\alpha_{\rm refl}$  & $0.20 \pm 0.01$ & $0.092 \pm 0.003$ & $0.006 \pm 0.004$ & $0.12 \pm 0.01$      & Reflection coefficient        \\

$\chi^2$ & 2.4 & 1.5 & 1.24 & 0.1 &  $\chi^2$ of the model \\
$P_{\rm}$ &$\rm1.8E-1$ $(1.4\sigma)$ & $\rm1.1E-3$ $(3.3\sigma)$ & $\rm 7.3E-1$ $(0.4\sigma)$& $\rm 9.8E-14$ $(7.4\sigma)$ & $\chi^2$-test confidence level \\ 
\\
\enddata
\end{deluxetable}
\capstarttrue

\clearpage

\section{The superrotation phase shift}
\begin{figure*} 
\centering
\resizebox{17cm}{14cm}
{
\includegraphics{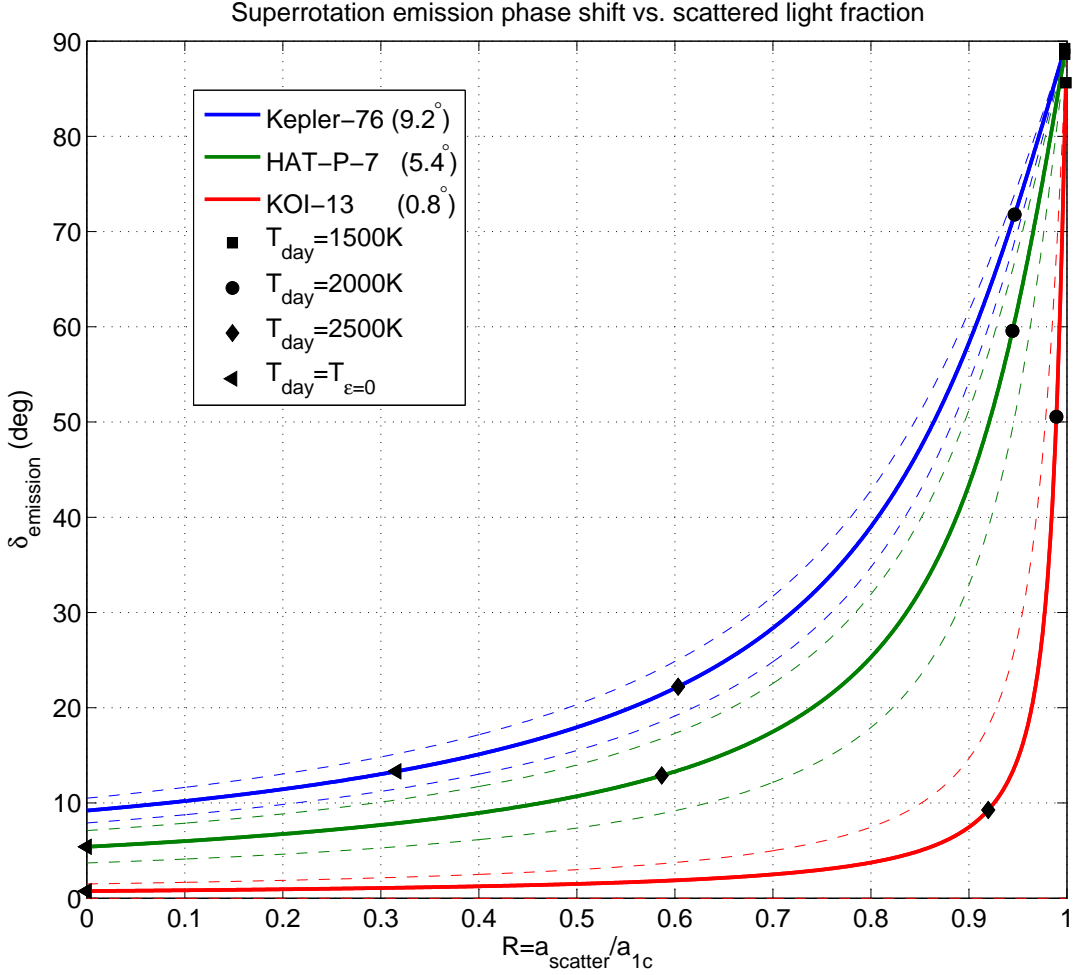}  
}
\caption{
Expected thermal emission phase-shift angle $\delta_{\rm emission}$ as a function of the ratio of scattered light to total unshifted reflected/emitted light $\mathcal{R}=a_{\rm scatter}/a_{1c}$ in the {\it Kepler} band for the three detections Kepler-76, HAT-P-7, and KOI-13. The dashed lines represent the $1\sigma$ range of the same color plot. The black markers show the expected system position on each plot for several dayside temperature values, assuming blackbody emission from the dayside and a dark nightside. $T_{\epsilon=0}$ is the no-albedo, no-redistribution, effective dayside temperature listed in Table~\ref{table:sys}.
}
\label{figure:angle}
\end{figure*}

The {\it Kepler}-band reflection/emission modulation is a combination of light scattered off the planet surface (reflection), together with radiation absorbed and later thermally reemitted (emission).
This point is important, as we expect superrotation to shift only the thermal reemission, while leaving the scattered-light component unshifted.
For the four systems the measured reflection/emission amplitude is smaller than, or similar to, the maximum thermal emission amplitude $A_{\rm ref,\epsilon=0}$ listed in Table~\ref{table:sys}, suggesting that the fraction of thermal emission in the visual {\it Kepler} light-curve phase modulation is probably significant.
Nevertheless, the superrotation phase shift that we derive in the {\it Kepler} band can serve only as a lower limit for the emission phase shift, while its actual value depends on the ratio 
\begin{equation} \label{eq:ratio}
\mathcal{R}=\frac{a_{\rm scatter}}{a_{1c}} \ ,
\end{equation}
where $a_{\rm scatter}$ is the scattered-light amplitude parameter and $a_{1c}$ is the derived total {\it unshifted} reflected/emitted amplitude, in the {\it Kepler} band. 
The $[0-1]$ range of $\mathcal{R}$, resulting from the $[0-a_{1c}]$ range of the $a_{\rm scatter}$ parameter, covers the entire range of emission-only to scatter-only planet irradiance, and any mixed emission/scattering in between, and is related to the thermal emission phase shift $\delta_{\rm emission}$ through
\begin{equation} \label{eq:deltaem}
\tan (\delta_{\rm emission})=\frac{\tan(\delta_{\rm sr})}{1-\mathcal{R}} \ .
\end{equation}
Figure~\ref{figure:angle} presents for the three detections Kepler-76, HAT-P-7, and KOI-13 the dependence of the thermal emission phase shift on $\mathcal{R}$, given the derived phase shifts in the {\it Kepler} band, and assuming geometric phase functions for both the scattered light and the emission modulations. The figure also marks the expected values for $\mathcal{R}$  at several dayside temperatures. The $\mathcal{R}$ value at each temperature was derived by estimating the emission amplitude as blackbody emission in the {\it Kepler} band from the dayside, assuming no heat redistribution, i.e., a dark nightside.

\clearpage

\section{Detailed comparison with previous studies} \label{sec:indiv}
In this section we compare our results with those of previous phase modulation studies and discuss specific features of each system.
\subsection{KOI-13}
\citet{shporer11}, \citet{mazeh12}, \citet{esteves13}, and \citet{angerhausen14} analyzed the {\it Kepler} light curve of KOI-13 and derived BEER amplitudes that are all within $1.4 \sigma$ of the amplitudes we report here, except for the beaming amplitude that was not detected by \citet{angerhausen14}. 
We also find a significant third harmonic amplitude of $7.4 \pm 0.5$ ppm with a phase shift of $1.1$ rad (see the KOI-13 residuals in Figure~\ref{figure:folds}), which is consistent with the KOI-13 third harmonic modulation detected by \citet{esteves13}. They suggest that this modulation may be a result of the gravity darkening of the fast-rotating host star KOI-13A.
For the planetary mass, \citet{mazeh12} and \citet{esteves13} report an inflated-beaming-based mass estimate, by $1.8 \sigma$ and $2.2 \sigma$ respectively, relative to the mass derived from the ellipsoidal amplitude, which are consistent with the findings of this work.
\citet{esteves13} and \cite{angerhausen14} also derived dayside brightness temperature for KOI-13b of $3706$ K and $3421$ K, respectively, which are close to our estimate of $3630$ K for $T_{\epsilon=0}$. 
Such dayside temperatures are consistent with an emission-only, zero-scattered-light phase function $(\mathcal{R}=0)$, resulting in an emission phase shift $\delta_{\rm emission}$ that is identical to the phase shift derived from the visual {\it Kepler} light curve, assuming a cold nightside (see Figure~\ref{figure:angle}).

\subsection{HAT-P-7}
As one of the most studied systems in the {\it Kepler} field, there are multiple studies of the HAT-P-7 {\it Kepler} light-curve phase modulations \citep{borucki09,welsh10,esteves13,angerhausen14}.
\citet{esteves13} analyzed the short-cadence Q0--Q14 {\it Kepler} light curve of HAT-P-7 and derived amplitudes that are within up to  $1.2 \sigma$ of the amplitudes we report here. They also derived an ellipsoidal-based planetary mass estimate that is consistent at the $1.6 \sigma$ level with the RV-derived mass, while reporting a significantly inflated-beaming-based mass estimate that is more than $13 \sigma$ larger than the RV-based estimate, a behavior that is consistent with the finding of this work. \citet{angerhausen14} analyzed the short-cadence Q0--Q15 {\it Kepler} light curve of HAT-P-7 but report amplitudes that differ by $1.2\sigma$--$3.0\sigma$ relative to the amplitudes we derive, a difference that may be a result of underestimated uncertainties. \citet{esteves13} also derived dayside brightness temperature of $2846$ K, which is close to our estimate of $2800$ K for $T_{\epsilon=0}$. 
Such temperatures are again consistent with a fully thermal, zero-scattered-light phase function $(\mathcal{R}=0)$ and a thermal-emission phase shift $\delta_{\rm emission}$ that is identical to the visual {\it Kepler} light-curve phase shift, assuming a cold nightside.  

\subsection{TrES-2}
 \citet{barclay12}, \citet{esteves13} and \cite{angerhausen14} used the short-cadence {\it Kepler} light curve of TrES-2 to derive its phase curve amplitudes, which are all within up to $1.2 \sigma$ of the amplitudes we report here. We note, though, that our amplitude uncertainties, which are derived from the quarter-to-quarter variations, are $2$--$4$ times larger than the uncertainties they report. For the planetary mass, \citet{barclay12} report about $2 \sigma$ difference between the mass derived from the ellipsoidal and the beaming amplitude, while \citet{esteves13} report consistent planetary mass derived from the two effects. 
Consistent with \citet{esteves13}, our analysis for TrES-2 shows no preference for a superrotation model, making our derived reflection/emission phase shift for this system consistent with zero.
\citet{esteves13} and \cite{angerhausen14} also derived dayside brightness temperatures of $1910$ K and $1947$ K, respectively, which are close to our estimate of $1880$ K for $T_{\epsilon=0}$. 
Such temperatures are consistent with zero scattered light $(\mathcal{R}=0)$, 
yielding thermal-emission phase shift $\delta_{\rm emission}$ that is consistent with zero, based on our analysis of the {\it Kepler} light curve.

\subsection{Kepler-76}
 In the planet discovery paper, \citet{faigler13} derived BEER amplitudes from the {\it Kepler} raw light curves of Q2--Q13 that are within $1.3 \sigma$ of the amplitudes derived here.
 \citet{angerhausen14} used the {\it Kepler} PDC light curves of Q0--Q15 to derive the phase curve amplitudes and the occultation depth of Kepler-76. Their derived amplitudes are again consistent within $1.3 \sigma$ with the amplitudes we report. They also obtained a planet brightness temperature of $2780$ K, which is close to our estimate of $2670$ K for $T_{\epsilon=0}$. 
Figure~\ref{figure:angle} shows that the derived $T_{\epsilon=0}$ yields minimum scattered-light ratio $\mathcal{R}$ of $0.32$ and minimum emission phase shift $\delta_{\rm emission}$ of $13{^\circ.}3$.

\section{Summary and discussion}
Several authors detected inconsistencies between planetary mass derived from the beaming amplitude and the one derived from the ellipsoidal amplitude for the four transiting hot Jupiters: KOI-13b, HAT-P-7b, TrES-2b, and Kepler-76b \citep{mazeh12,shporer11,esteves13,barclay12,faigler13}. In addition, RV measurements, available for HAT-P-7, TrES-2, and Kepler-76 \citep{winn09,odonovan06,faigler13}, show  spectroscopic RV amplitudes that are significantly smaller than the beaming-derived RV ones, pointing to inflated beaming amplitudes.
In their discovery paper, \citet{faigler13} suggested that the inconsistency of the Kepler-76 beaming amplitude can be explained by a phase shift of the reflection/emission modulation due to the hot Jupiter superrotation phenomenon predicted by \citet{showman02} and later observed by \citet{knutson07,knutson09,knutson12} in the infrared. Here we extend and test this model also for KOI-13b, HAT-P-7b, and TrES-2b.

To allow the BEER model to account for superrotation, we developed analytic approximations for the amplitudes of the first two harmonics of the BEER modulation of a hot Jupiter system, assuming
\begin{itemize}
\item a circular orbit;
\item that planetary mass is negligible relative to the star mass;
\item a first-order approximation for the ellipsoidal variation;
\item a superrotation-induced phase-shifted Lambertian or geometric reflection/emission phase function.
\end{itemize}



For Kepler-76 and HAT-P-7 $\chi^2$-tests show that the Lambert superrotation BEER model yields a better fit to the data and is highly preferred over the unshifted null model, while for KOI-13 it is preferable only at the $1.4 \sigma$ level. For TrES-2 we find no preference for the superrotation model. Nevertheless, for all four systems the planet mass estimate derived from the Lambert superrotation BEER model is highly consistent with the planetary mass derived or constrained by RV studies, suggesting that the Lambertian superrotation model yields a good photometric estimate for the planet mass, given a good stellar model.

Initially, the phase-shifted emission modulation was identified in the {\it Kepler} band owing to its ``leakage'' into the $a_{1s}$ coefficient, resulting in an apparently inflated beaming amplitude. It is interesting to check the dependence of this ``leakage'' phenomenon on the planetary parameters of the system. Using the relations in Table~\ref{table:amps}, the relative addition to the $A_{\rm beam}$ amplitude, due to phase-shifted emission, is
\begin{equation} \label{eq:prop}
 \frac{A_{\rm ref}\sin\delta_{\rm sr}}{A_{\rm beam}} \propto \frac{R_{\rm p}^2}{M_{\rm p} P_{\rm orb}} \,
\end{equation}
while the right-hand side of the equation results from Equation~(\ref{eq:amps}). Considering that over the secondary-mass range of $1$--$100 M_{\rm Jup}$, covering Jupiters to late M dwarfs, the radius remains almost unchanged at about  $1 R_{\rm Jup}$, the dependence above suggests that the relative inflation of the beaming amplitude is at maximum at the $1 M_{\rm Jup}$ end of the range. Adding to that the $P_{\rm orb}$ dependence, we conclude that close-in, hot Jupiters are expected to show the most apparent inflated beaming amplitude. It is then not a surprise that this phenomenon was initially discovered in hot Jupiters and is in agreement with the phase-shifted reflection/emission modulations of Kepler-76, HAT-P-7, and KOI-13 reported by this study.

Detailed phase curve studies, such as the one we present here and alike \citep[e.g.,][]{esteves14}, open the opportunity to estimate the mass and investigate the atmospheric properties of multiple close-in exoplanets, through analysis of the precise photometric light curve produced by space telescopes like CoRoT and {\it Kepler}. Such analyses of photometric light curves of future missions, like TESS and PLATO, can not only discover nontransiting stellar binaries and exoplanets \citep{faigler12,faigler13} but also provide consistent planetary mass and density estimates of transiting exoplanets, and even identify close-in planets with unique, or nonordinary, phase-curve behavior, as targets for spectroscopic and IR follow-up. These phase-curve studies can serve as a highly efficient filter for focusing the research on transiting exoplanet systems with the most intriguing mass, radius, density, and phase curve features, so that future IR observations and transmission-spectroscopy resources are efficiently assigned to systems that are most valuable for our understanding of planetary atmospheres.


\acknowledgments
We are indebted to Shay Zucker for numerous helpful discussions. 
We thank the anonymous referee for his highly valuable remarks and suggestions.
The research leading to these results has received funding from the European Research Council under the EU's Seventh Framework Programme (FP7/(2007-2013)/ ERC Grant Agreement No.~291352).
This research was supported by the Israel Science Foundation (grant No.~1423/11) and the Israeli Centers Of Research Excellence (I-CORE, grant No.~1829/12).
We feel deeply indebted to the team of the {\it Kepler} mission, which enabled us to search and analyze their unprecedentedly accurate photometric data.
All the photometric data presented in this paper were obtained from the 
Multimission Archive at the Space Telescope Science Institute (MAST). 
STScI is operated by the Association of Universities for Research in Astronomy, Inc., under NASA contract NAS5-26555. Support for MAST for non-{\it HST} data is provided by the NASA Office of Space Science via grant NNX09AF08G and by other grants and contracts.

\end{document}